\documentclass[aps,prl,twocolumn,groupedaddress,showkeys,showpacs]{revtex4}

\usepackage{graphicx}
\usepackage{amssymb}
\usepackage{epstopdf}
\begin{document}


\title{An artificial dynamic heterogeneity in a glass-former below  Tg\\}


\author{V. Teboul$^{1}$, M. Saiddine$^{1}$, J-M. Nunzi$^{2}$}
\affiliation{ 
1. Laboratoire des Propri\'et\'es Optiques des Mat\'eriaux et Applications, FRE CNRS 2988, Universit\'e d\'\ Angers, 2 Bd Lavoisier, F-49045 Angers, France \\
 2. Departments of chemistry and physics at Queen\'\ s University, Kingston K7L 3N6, Ontario, Canada\\}


\date{\today}

\begin{abstract}
We report the first large scale molecular dynamics simulations of the effect of the photoisomerization of probe molecules on the dynamics of a glassy molecular material. 
We show that the photoisomerization of the probe molecules induces the creation of an artificial dynamical heterogeneity inside the matrix. We then study the modification of the dynamics and find, in agreement with recent experiments, an important increase of the diffusion leading to liquid like diffusion below the glass transition temperature. 
We find that the photoisomerization process controls the heterogeneity and the non-Gaussian parameter of the material, leading to extremely rapid variations of these quantities. 
\end{abstract}


\maketitle


A central question in the field of the glass-transition is the relation between dynamic heterogeneities and the relaxation dynamics of supercooled liquids and glassy materials \cite{rev1,rev2,adam}. In order to shed some light on this relation, we investigate various methods to modify the dynamic heterogeneities, and then to observe the resulting effects in the relaxation dynamics. Because it induces a cutoff in the correlation length of any cooperative mechanism present in the material, confinement has been seen as a promising direction of study. The liquid may be confined inside pores, micelles, or nanoparticles may be dispersed inside the liquid. However in most cases the surface effects predominate in the origin of the dynamical properties of confined supercooled liquids. For example, depending on the surface properties of the confining medium the dynamics is slowed down or accelerated\cite{sch1,krakov,us1,horb}. \\
\\
In this letter we present a different and more direct method to modify the dynamic heterogeneities inside a supercooled or a glassy material. 
We show that the photoisomerization of probe molecules inside the glassy or supercooled matrix induces the creation of an artificial dynamical heterogeneity inside the matrix. We then study the modification of the dynamics and find,  in accordance with recent experimental results\cite{natmat}, an important increase of the diffusion. We find a strong modification of the plateau of the mean square displacement (MSD), suggesting induced cage breaking process as a cause for the acceleration of the dynamics.
We find that the photoisomerization process controls the heterogeneity and the Non-Gaussian parameter (NGP) of the material, leading to extremely rapid variations of these quantities. 
We simulate the photoisomerization of dispersed red ($DR1$) molecules (probes) inside a matrix of methylmethacrylate ($MMA$) molecules ($C_{5}O_{2}H_{8}$). We study the effect of the probe isomerization on the matrix dynamics and on the dynamical heterogeneity inside the matrix.
Under continuous holographic exposure of thin polymer films doped with azo-dyes Rochon et al. and Kim et al.\cite{orig1,orig2} observed the formation of surface relief gratings (SRG) corresponding to massive photoinduced mass transport\cite{revazo1,revazo2,revazo3,natmat} well below the glass transition temperature of the material. Various theories have been proposed to explain this effect\cite{theo1,theo2,theo3,theo4,theo5}. Recently SRG have also been observed in molecular glasses\cite{molecular1,molecular2,molecular3} suggesting universality in the mechanism of SRG formation. 
An increase of the local diffusion has been suggested \cite{theo1,theo4} as a possible cause for SRG formation but is still a matter of controversy.\\
\\
The present simulations were carried out for a system of 700 $MMA$ molecules and 1 or 7 $DR1$ molecules. Our simulations use the Gear algorithm with the quaternion method \cite{allen} to solve the equations of motions with the following potential functions \cite{pot1,pot2}. The time step was chosen equal to $0.5$ $10^{-15} s$. The density was set constant in our simulations at $1.19 g/cm^{3}$. The temperature was controlled using a Berendsen thermostat \cite{beren}. The observed maximum fluctuations were of ${+/-}$ $ 8 K$. 
The $DR1$ molecules are in some simulations treated as rigid bodies in the $trans$ or $cis$ isomer in order to study the dynamics without photoisomerization. In other simulations, in order to study the effect of photoisomerization, the $DR1$ molecules change periodically and continuously from the $trans$ to the $cis$ isomer and then from the $cis$ to the $trans$ isomer. In this case the $DR1$ molecule changes from $trans$ to $cis$ uniformly in a time $t_{1}=300 fs$. This value is typical of the isomerization process \cite{fermeture}. Then it stays in the $cis$ configuration during a time $t_{2}$, changes uniformly from $cis$ to $trans$ in a time $t_{1}$ and stays in the $trans$ configuration during a time $t_{2}$. Results presented in this letter correspond to $t_{2}=t_{1}$ then a short waiting time value. We have then tested the effect of the time variation of $t_{1}$ and $t_{2}$  on the matrix dynamics \cite{prep}. We find\cite{prep} that the dependence is roughly linear with time variations. Typical experimental light exposure leads to a value of $1ms$ between isomerizations. Taking also into account the weak DR1 concentration of our simulations, we estimate that our diffusion coefficient must be multiplied by a rough factor $10^{-7}$ if compared with typical SRG experiments. 
We define the mobility $\displaystyle {\mu_{i,t_{0}}(t)}$ of the atom  $i$  at time $t_{0}$ with a characteristic time $t$, as the displacement of the atom $i$ during the time $t$\cite{het3}:
$$\displaystyle {\mu_{i,t_{0}}(t)= |{\bf{r}}_{i}(t+t_{0})-{\bf{r}}_{i}(t_{0})| }\eqno [1] $$
  
The dynamical heterogeneity are characterized \cite{lj,y,vvh,polymer}  by the transient aggregation of the molecules of highest mobility in the liquid. This concerns typically the 5 percent most mobile molecules.   Similarly a transient aggregation of the molecules of low mobility is observed \cite{het2,het3} but with different characteristic times. 
These aggregations are the fingerprints of the cooperativity of motions inside the supercooled liquid and disappear around the melting temperature $T_{m}$\cite{lj,y,vvh,polymer}. 
In order to quantify the aggregation of the most mobile molecules we define the function:
$$I^{+}(t)= \int_0^{R_{c}} (g_{m-m,t}(r)/g(r) - 1) 4 \pi r^{2} dr \eqno [2] $$
 
 and similarly for the least mobile molecules:
 $$I^{-}(t)= \int_0^{R_{c}} (g_{lm-lm,t}(r)/g(r) - 1) 4 \pi r^{2} dr \eqno [3] $$
 
Where $g_{m-m,t}(r)$ is the radial distribution function (RDF) between the most mobile matrix molecules. And $g_{lm-lm,t}(r)$ is the RDF between the least mobile molecules, while $g(r)$ is the RDF between random matrix molecules.
The time $t$ appearing in these functions is the time used in the definition (1) of the mobility ${\mu_{i}(t)}$. This time is then used in the selection of the most or the least mobile molecules. The cut-off  $R_{c} =6$ \AA$ $  has been chosen to reduce noise effect. 
For text clarity we will name functions $I^{+/-}(t)$ intensity of the aggregation.
When the time t corresponds to the phenomena of physical importance for the motion of molecules, the aggregation appears. The aggregation of mobile molecules then appears usually \cite{lj} for $t=t^{*}$ which corresponds to the maximum value of the NGP. This time scale corresponds roughly to the mean time used by a molecule to escape the cage created by its neighbors. For the least mobile molecules the characteristic time is usually larger and this difference increases when the temperature decreases. This effect has been related to the breaking of the Stokes Einstein relation \cite{rev1,einstein}. These behavior have been observed in a number of supercooled liquids \cite{rev1,lj,y,vvh,polymer}. When the photo-isomerization is switched off, we observe this behavior in the $MMA$ + $DR1$ material investigated here \cite{prep}.\\
\\
\begin{figure}
\includegraphics{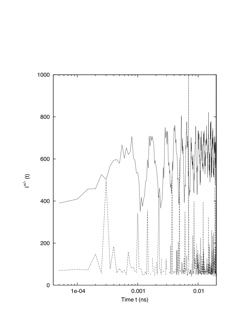}
 \caption{ Intensity of the mobiles heterogeneities   with $R_{c} = 6$ \AA , and of the slow heterogeneities, for the centers of masses of the matrix molecules. The photoisomerization is switched on. The temperature is 180 K, then well below the glass transition temperature ($T_{g}=400K$). The simulation box contains one $DR1$ molecule and 700 $MMA$ molecules.}
\end{figure}

When the photo-isomerization is switched on, the behavior of the dynamic heterogeneities changes totally. Figure 1 shows the intensity versus time $ I^{+}(t)$ (eq. 2) of the aggregation of the most mobile matrix molecules in this case
. The temperature of this simulation ($180K$) is largely below the glass-transition temperature. At this temperature the time scale of the heterogeneity when the isomerization is switched off is larger than $1 ns$ \cite{prep}. The time scales in figure 1 are much shorter. The intensity $I^{+}(t)$ oscillates rapidly between maxima and minima. The most mobile matrix molecules aggregate when the $DR1$ probe molecule undergoes a $trans$-$cis$ isomerization. And these aggregates disappear partly when the probe molecule undergoes a $cis$-$trans$ isomerization. 
To our knowledge the time scale of the heterogeneity observed here is much shorter than any time scale reported to date. This time scale corresponds exactly to the time scale introduced in the simulation for the isomerization. This result shows that the heterogeneity in the matrix is here controlled by the isomerization of the probe molecule. 
Figure 1 also shows the intensity $I^{-}(t)$ of the aggregation of the least mobile molecules. $ I^{-}(t)$ displays peaks at periodic times. The slow heterogeneity is maximal when the $DR1$ is in the $trans$ isomer. The slow heterogeneity then disappears completely when the $DR1$ undergoes a $trans$-$cis$ isomerization. We observe in figure 1 that the intensities $ I^{+}(t)$ and $ I^{-}(t)$, corresponding to the aggregation of mobile and slow molecules, follow an opposite behavior.
When the $DR1$ undergoes a $cis$-$trans$ isomerization the slow heterogeneity appears and increases sharply to a maximum value. At the same time the mobile heterogeneity decreases sharply. Then when the $DR1$ undergoes a $trans$-$cis$ isomerization the opposite behavior appears. The slow heterogeneity collapses and the mobile heterogeneity increases to a maximum value. We observe here for the first time, a total time separation between the mobile and slow heterogeneity. The time evolution of the heterogeneity coincides with the time evolution of the isomerization process. When the photo-isomerization is switched off the heterogeneous behavior of this material at this temperature corresponds to characteristic times larger than $1 ns$. No heterogeneous behavior is then observed on the time scales of figure 1 when the photo-isomerization process is switched off. In the material studied here, the photo-isomerization of the $DR1$ then controls totally the slow and mobile heterogeneities. These results show the possibility of controlling the heterogeneity inside a material with an external input.\\
\\
\begin{figure}
\includegraphics{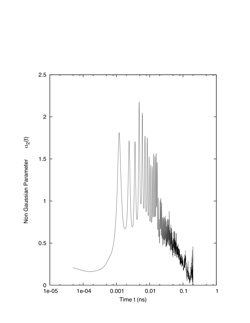}
\caption{Non Gaussian parameter $\alpha_{2}(t)$ of the centers of masses of the matrix molecules when the photoisomerization is switched on. The temperature is 180K. The simulation box contains one $DR1$ molecule and 700 $MMA$ molecules.}
\end{figure}

We observe in Figure 2, peaks and dips superimposed on a main smooth curve. The main smooth curve has an usual shape for $\alpha_{2}(t)$. But the time scale correspond to a viscous liquid at a much higher temperature. The characteristic time that corresponds to the maximum of the main smooth curve is $t^{*}_{1} = 8$ $ ps$. While at the same temperature the characteristic time $t^{*}$ of $\alpha_{2}(t)$ is higher than $1 ns$ when the isomerization is switched off. The characteristic time of the main curve $t^{*}_{1}$ is then very unusual.
Superimposed on this main curve, peaks and dips appear with the characteristic times of the $trans$ and $cis$ lifetimes used in our simulation. The time scales of these peaks and dips are very short. The correspondence between these time scales and the $trans$ and $cis$ life times show that this behavior is also controlled by the isomerization process. The $DR1$ isomerization controls then the NGP value, which is an important characteristic of the non-Markov dynamics.
\begin{figure}
\includegraphics{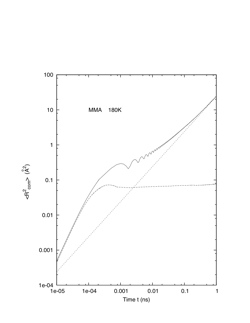}
\caption{Mean square displacement $<r^{2}(t)>$ versus time at a temperature of 180K, when the photoisomerization is switched on (upper curve) or off (lower curve). The simulation box contains 7 $DR1$ molecule and 700 $MMA$ molecules.}
\end{figure}
Figure 3 shows the MSD of the matrix molecules when the $DR1$ undergoes or not photo-isomerization. At this temperature when there is no isomerization,
 the motion is stopped on the time scale of the simulation. This is shown on the lower curve of Figure 3. The plateau of the MSD begins around $1 ps$ and continues up to the larger times of the simulation. However when the isomerization is switched on, the plateau of the MSD disappears and is replaced by oscillations.  Figure 3 shows that for large times the MSD follows the Einstein law \cite{hansen} for Brownian motion (dashed line).  For large times we then observe a liquid-like diffusion law. The diffusion coefficients $D$ calculated from this law are displayed in Figure 4. These diffusion coefficients are typical of  viscous liquids and not of glassy materials. We then observe in Figure 3 a liquid-like diffusion mechanism at a temperature below $T_{g}$. 
This strange comportment is validated by recent experimental data \cite{natmat}. 
The oscillations in the MSD correspond to the $cis$-$trans$ isomerization process. The MSD is maximum for the $trans$-$cis$ isomerization and minimum for the $cis$-$trans$ isomerization. 
The comportment of the matrix for $trans$-$cis$ isomerization approaches the comportment of a liquid. The NGP decreases showing that the dynamics tends to a Markov dynamics; The slow heterogeneity disappears and the diffusion coefficient $D$ approaches viscous liquid values. However the presence of mobile heterogeneity shows that the matrix does not behave totally as a normal liquid.
Amazingly, the comportment of the matrix for $cis$-$trans$ isomerization is also not the comportment of a glassy material. Approaching the comportment of a glass, the slow heterogeneity increases, the NGP increases and the MSD decreases. However the MSD is higher than the plateau of the glass, as seen in figure 3. Another striking difference is that the characteristic time of the slow heterogeneity is not increasing with temperature decrease. The characteristic times for slow and mobile heterogeneity and NGP are here imposed by the isomerization characteristic times.
When the isomerization $trans$-$cis$ is switched on, it imposes a cage breaking process. Figure 3 shows that the plateau, which is a characteristic of the cage effect, disappears. This cage breaking process follows then the characteristic times of the isomerization. 
When the isomerization is switched off,  the diffusion coefficient of the matrix molecules, doped with  $trans$ or  $cis$ $DR1$ isomers, are very similar\cite{prep}. 
 The $trans$ $DR1$ isomer has been chosen in the figures.
 \begin{figure}
\includegraphics{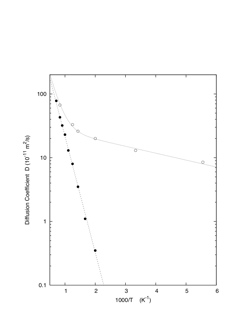}
 \caption{Diffusion coefficient for the matrix, as a function of the inverse of temperature, when the photoisomerization is switched on: Empty circles; or off: Full circles. The dashed line fit (photoisomerization switched off) corresponds to an Arrhenius dependence $D=D_{1}exp(-E_{a1}/T)$ with $E_{a1}= 4000 K$.   When the isomerization mechanism is switched on, the diffusion coefficient increases. The dotted line fit (photoisomerization switched on) follows a law of the form: $D= D_{1}exp(-E_{a1}/T) + D_{2}exp(-E_{a2}/T)$ with $E_{a2}=250K$.  This result suggests that the isomerization adds a new Arrhenius process to the original dynamics of the matrix. We estimate that $D$ must be multiplied by a rough factor $10^{-7}$ if compared with typical SRG experimental conditions. }
\end{figure}

Figure 4 shows the diffusion coefficient of the matrix molecules versus temperature when the isomerization is switched on or off. 
When the isomerization is switched off we have an Arrhenius evolution of the diffusion coefficient. 
We observe a typical Arrhenius dependence $D=D_{1}exp(-E_{a1}/T)$ with $E_{a1}= 4000 K$.  The diffusion coefficients in Figure 4 may also be fitted with an MCT law:
 $D=D_{0}(T-Tc)^{\gamma}$ with $Tc=447.2 K$  and $\gamma=2.3$.  The MCT fit leads to an extrapolation of $T_{g}$ around $400K$.
 When the isomerization mechanism is switched on, the diffusion coefficient increases. We obtain the sum of two Arrhenius mechanisms.
The diffusion coefficient follows a law of the form: $D= D_{1}exp(-E_{a1}/T) + D_{2}exp(-E_{a2}/T)$ with $E_{a2}=250K$.  This result suggests that the isomerization adds a new Arrhenius process to the original dynamics of the matrix. \\
 \\
In conclusion, our study shows that the isomerization of probe molecules in glassy materials creates a dynamical heterogeneity inside the material.
This heterogeneity is controlled by the isomerization process. 
This result suggests a new method to study the effect of the heterogeneity on the material dynamics, using the photoisomerization of probe molecules to control the heterogeneity.
To our knowledge, this is a unique case where the slow heterogeneity and the mobile heterogeneity appear at different times and thus are not present together. This result shows that the two kinds of heterogeneity are not directly related. i.e. that the slow heterogeneity is not the cause of the presence of the mobile heterogeneity in supercooled liquids.
These results suggest a possible control of the dynamical properties of amorphous materials and of the glass-transition mechanism via the creation of heterogeneity which may lead to a new route for cryopreservation of biomaterials.
The huge effect observed on the dynamical properties of the material suggests a strong relation between the presence of dynamical heterogeneity and the dynamical properties.
This study also shows that both kinds of aggregations (slow and mobile) have a different effect on the dynamics. This result has to be taken into account in the Adam et Gibbs\cite{adam} interpretations of the glass-transition when mobile heterogeneity are used as a measure of the configurational entropy.\\

\end{document}